\documentclass{article}

\usepackage{PRIMEarxiv}

\usepackage[utf8]{inputenc} 
\usepackage[T1]{fontenc}    
\usepackage{hyperref}       
\usepackage{url}            
\usepackage{booktabs}       
\usepackage{amsfonts}       
\usepackage{nicefrac}       
\usepackage{microtype}      
\usepackage{lipsum}
\usepackage{fancyhdr}       
\usepackage{graphicx}       
\usepackage{float}
\usepackage{subcaption}
\usepackage{amsmath}
\usepackage{tabularx}
\usepackage{titlesec}
\titleformat{\section}[block]{\normalfont\Large\bfseries}{\thesection.}{1em}{}

\title{\textbf{Particle Multi-Axis Transformer for Jet Tagging}}

\author{
  \textbf{Muhammad Usman}\textsuperscript{4},
  \textbf{M Husnain Shahid}\textsuperscript{4}, 
 \textbf{Maheen Ejaz}\textsuperscript{4},
  \textbf{Ummay Hani}\textsuperscript{4}, 
  \textbf{Nayab Fatima}\textsuperscript{4},\\
 \textbf{Abdul Rehman Khan}\textsuperscript{1},
  \textbf{Asifullah Khan}\textsuperscript{1,2,3*}, 
  \textbf{Nasir Majid Mirza}\textsuperscript{4}
  \\
  \textsuperscript{1} Pattern Recognition Lab, Department of Computer \& Information Science, \\Pakistan Institute of Engineering \& Applied Sciences, Nilore, Islamabad 45650, Pakistan \\
  \textsuperscript{2} PIEAS Artificial Intelligence Center (PAIC), \\Pakistan Institute of Engineering \& Applied Sciences, Nilore, Islamabad 45650, Pakistan \\
  \textsuperscript{3}Center for Mathematical Sciences, \\Pakistan Institute of Engineering \& Applied Sciences, Nilore, Islamabad 45650, Pakistan \\
  \textsuperscript{4}Department of Physics \& Applied Mathematics, \\Pakistan Institute of Engineering \& Applied Sciences, Nilore, Islamabad 45650, Pakistan \\
  \texttt{Corresponding Author: \textsuperscript{*}Asifullah Khan, asif@pieas.edu.pk}}

\begin{document}
\maketitle

\begin{abstract}
Jet tagging is an essential categorization problem in high energy physics. In recent times, Deep Learning has not only risen to the challenge of jet tagging but also significantly improved its performance. In this article, we proposed an idea of a new architecture, Particle Multi-Axis transformer (ParMAT) which is a modified version of Particle transformer (ParT). ParMAT contains local and global spatial interactions within a single unit which improves its ability to handle various input lengths. We trained our model on JETCLASS, a publicly available large dataset that contains 100M jets of 10 different classes of particles. By integrating a parallel attention mechanism and pairwise interactions of particles in the attention mechanism, ParMAT achieves robustness and higher accuracy over the ParT and ParticleNet. The scalability of the model to huge datasets and its ability to automatically extract essential features demonstrate its potential for enhancing jet tagging.
\end{abstract}

\keywords{Jet Tagging \and Deep Learning \and Transformer \and Multi-Axis-Attention}

\section{Introduction}

The largest and most robust particle accelerator in the world is the Large Hadron Collider (LHC), which is located at CERN. It is made up of a circular tube that is 27 km long and situated between 50 and 175 meters below the surface. New particles are created inside the collider when beams collide with each other. Reaching an energy level of 14 TeV in the center of the mass reference frame is the main goal of the Large Hadron Collider. The breakthrough of Higgs Boson in 2012 was one of its most important accomplishments. Nine experiments are carried out at the LHC which are facilitated by collisions within the accelerator. A variety of particle detectors are used in this research. Four of these are large-scale detectors.Two of them are general-purpose detectors such as ATLAS (A toroidal LHC Apparatus) and CMS (Compact Muon Solenoid), and two are research-focused detectors such as LHC-b and ALICE (A Large Ion Collider Experiment). These huge detectors are positioned beneath the LHC in vast caverns. The Large Hadron Collider Forward (LHCf), the Forward Search Experiment (FASER), the Scattering and Neutrino Detector at the LHC (SND@LHC), the MoEDAL (Monopole and Exotics Detector at the LHC), and TOTEM (Total, elastic and diffractive cross-section measurement) are the five small detectors. Among these, the ATLAS stands as one of the largest detector, measuring 46m in length, 25m in height, and 25m in breadth. All detectors transfer some part of or all of the radiation energy to the detector mass. The form in which energy is converted depends on the type of detector. A gaseous detector directly collects ionization electrons to form an electric signal. In Scintillation detectors, both excitation and ionization induce molecular transition which results in the emission of light. These detectors are capable of producing a usable signal for a given type of radiation and energy. The Solenoid controls the path of particles inside the detector. Neutrino remains invisible in the detector.

Machine learning has enhanced the possibility of discovering new fundamentals of nature and  has transformed the methodology for analyzing large-scale data samples in High Energy Physics. At the LHC, proton beams are propelled to almost the velocity of light and then smash at 40 million times per second (40 MHz).  These energetic proton-proton smashes yield new, short-lived particles that subsequently decay, producing a concentrated spray of outgoing particles referred to as jets. Particle detectors such as CMS or ATLAS are tasked with detecting these jets. Jet Tagging is to identify and classify jets that are produced by particles such as quark and gluons\cite{he2023quark}, the W or Z boson, or the Higgs boson\cite{spira1995higgs}. It is a critical task in High energy physics because the particles that creates jets radiates and these radiated particles produces more outgoing particles. It will be difficult to identify the original particle because these radiations distorts their original features. Traditional approaches to jet tagging depend on handcrafted features inspired by concept of Quantum Chromodynamics(QCD). QCD is a  theoretical framework that describes how particles moves within a jet. Different methods have been adopted so far for the identification of jets. The dominant representation describe a jet as a cloud of particles, characterized by an unorganized and variable sized collection of particles that are released from detectors.

Deep learning has transformed Jet tagging. We harnessed the idea of a new architecture, Particle Multi-axis Transformer (ParMAT) for the classification tasks of jets, which fits with the JETCLASS data sets. It simplifies the complex attention mechanism used in the Multi-Axis Vision Transformer(MaxViT)\cite{tu2022maxvit}. We used JETCLASS dataset for training this model. It is large and detailed dataset consisting of 100M jets for training. It has 10 classes and it also includes the jets that have not been explored yet for tagging. The resulting Particle Multi-axis transformer(ParMAT) achieves higher performance than particle transformer(ParT), PFN, P-CNN and ParticleNet. Training our model led to notable increase in accuracy.

\section{Related Work}

\textbf{Deep Learning for Jet tagging:} Deep learning techniques like CNN and Transformers have shown promising results in different domains\cite{khan2024venturing}, and these techniques are also gaining popularity for jet tagging as it has significantly improved particle identification in recent years\cite{qu2022particle}\cite{komiske2019energy}\cite{drugakov2020identification}\cite{qu2020jet}. The efficiency and effectiveness of DL techniques on jet physics has improved with the advancement of two factors.

\begin{itemize}
\item \textbf{Jet Representation:}  Previously jets were represented as images\cite{de2016jet}, graphs\cite{henrion2017neural}, sequences\cite{guest2016jet} and trees\cite{henrion2017neural} which worked well with corresponding architectures like graph neural networks, 2D Convolutional Neural networks and recurrent networks. In the recent times, jet as particle cloud\cite{komiske2019pythia8}\cite{qu2020jet} has been introduced, comparable to point clouds, which view a jet as an invariant set of particles under permutations. Based on this representation, architectures like Energy Flow Networks\cite{komiske2019pythia8}, Graph Attention based Point Neural Network\cite{chen2021gapointnet}, Point Cloud Transformer\cite{guo2021pct} were introduced. Among these, the new Particle Net remained the predominant DL architecture.
\item \textbf{Jet Tagging Datasets:} Simulated jet tagging datasets has proved to be more convenient than the real ones, For this investigation, a number of datasets have been released so far. These include Top quark\cite{kasieczka2019machine}, Quark gluon\cite{komiske2019pythia8}, and Higgs Bosons\cite{cms2019sample} tagging datasets. These along with JetNet\cite{kansal2021particle} and multiclass dataset\cite{moreno2020jedi} contain information about kinematics and particle identification only. Architectures like 1D CNNs, RNN used these datasets for studies. However, uptil now, the recently introduced JetClass has the most information and maximum number of different types of jets . This dataset contains immense information; thus it needs relatively bigger models like the recent Particle transformer.
\end{itemize}

Deep learning powered jet tagging techniques have become integral to data analysis at the LHC, notably through innovations like the DeepAK8 algorithm developed by the CMS Collaboration\cite{drugakov2020identification}. This algorithm, employing a 1D CNN\cite{asam2022iot} based on the ResNet architecture\cite{he2016deep}, effectively identifies jets arising from top quarks or bosons like the Higgs, W, or Z. Such advancements have markedly heightened the potential for discovering new heavy particles, as evidenced by CMS publications in 2021\cite{belforte2021search} and 2022\cite{qu2022particle}. Additionally, ParticleNet, another deep-learning tool utilized by CMS, enabled groundbreaking achievements, including the first observation of Z boson decay into charm quarks and setting stringent constraints on Higgs decay to charm quarks\cite{qu2022particle}. Furthermore, ParticleNet's application in probing the quartic Higgs and vector bosons interaction has led to the indirect confirmation of its existence. These strides underscore the crucial role of improved jet tagging methods in advancing our knowledge of elementary particles, the fundamental constituents of the universe. 

\textbf{Transformers:} In recent years, transformer models have achieved remarkable success, initially in natural language processing and later expanding into computer vision. The original Transformer architecture\cite{vaswani2017attention}, along with its derivatives like BERT\cite{devlin2018bert}, ViT\cite{dosovitskiy2020image}, and Swin-Transformer\cite{liu2021swin}, has consistently set new performance benchmarks across diverse tasks, showcasing the versatility of Transformers. Beyond traditional domains, Transformers, driven by their attention mechanism, have proven adept at addressing fundamental scientific challenges. Notably, AlphaFold2\cite{jumper2021highly}, a groundbreaking model in protein structure prediction, leverages the attention mechanism, enhancing its predictive accuracy\cite{hayat2012mem}. Specifically, interpretability of the model is improved by incorporating a pair bias into the self-attention mechanism, which is obtained from pairwise features.. These advancements highlight the transformative impact of Transformer architectures across both conventional and scientific domains, highlights their universal applicability and effectiveness.

\textbf{Transformer in vision:} In computer vision, the innovative perspective of ViT, which takes image patches as visual words, has sparked an intense surge of research interest in visual Transformers. Vit and its derivatives are commonly utilized in image classification\cite{khan2023survey}, segmentation\cite{khan2023recent}\cite{khan2023maxvit}, object detection and visual recognition tasks. The variants like Deit\cite{touvron2021training}, Beit\cite{bao2021beit}, and Swin\cite{liu2021swin} operate by iteratively focusing on pertinent segments of the input sequences, updating representations via self-attention mechanisms to identify complex relationships and patterns.  Improved locality, pyramidal designs, training techniques, sparse attention, and improved locality are some of the more recent efforts aimed at improving model and data efficiency. In this domain, Multi axis Vision transformer(MaxVit) has been introduced in 2022\cite{tu2022maxvit} and RSIR Transformer\cite{zhang2023rsir} in 2023 which show superior performances as compared to other common vision transformers.

\section{Dataset}

We used the JETCLASS dataset to train our proposed Particle Multi-axis Transformer for the jet tagging task. JETCLASS is a large-scale, simulated dataset for CMS detector, proposed and generated by Hulin Qu\footnote{CERN, Geneva, Switzerland}, Congqiao Li\textsuperscript{2} and Sitian Qian\footnote{School of Physics, Peking University, Beijing, China}\cite{qu_particle_2024}, to facilitate deep learning research in jet tagging. This dataset includes 10 types of jets, including signal jets from the decay of $WZ$ Bosons, top quarks ($t$), Higgs ($H$) bosons and background signal from quark-gluon pairs ($q/g$). Since, in particle collision experiments, a particle has more than one mode of decay, which means each decay mode results in a different type of jet. So, this dataset considers 5 decay modes of Higgs, two decay modes of top quarks, one decay mode for W and Z bosons each, and background jets from quark-gluon plasma, resulting in 10 types of jets.

\begin{figure}
    \centering
    \captionsetup{justification=centering}
    \includegraphics[width=0.5\linewidth]{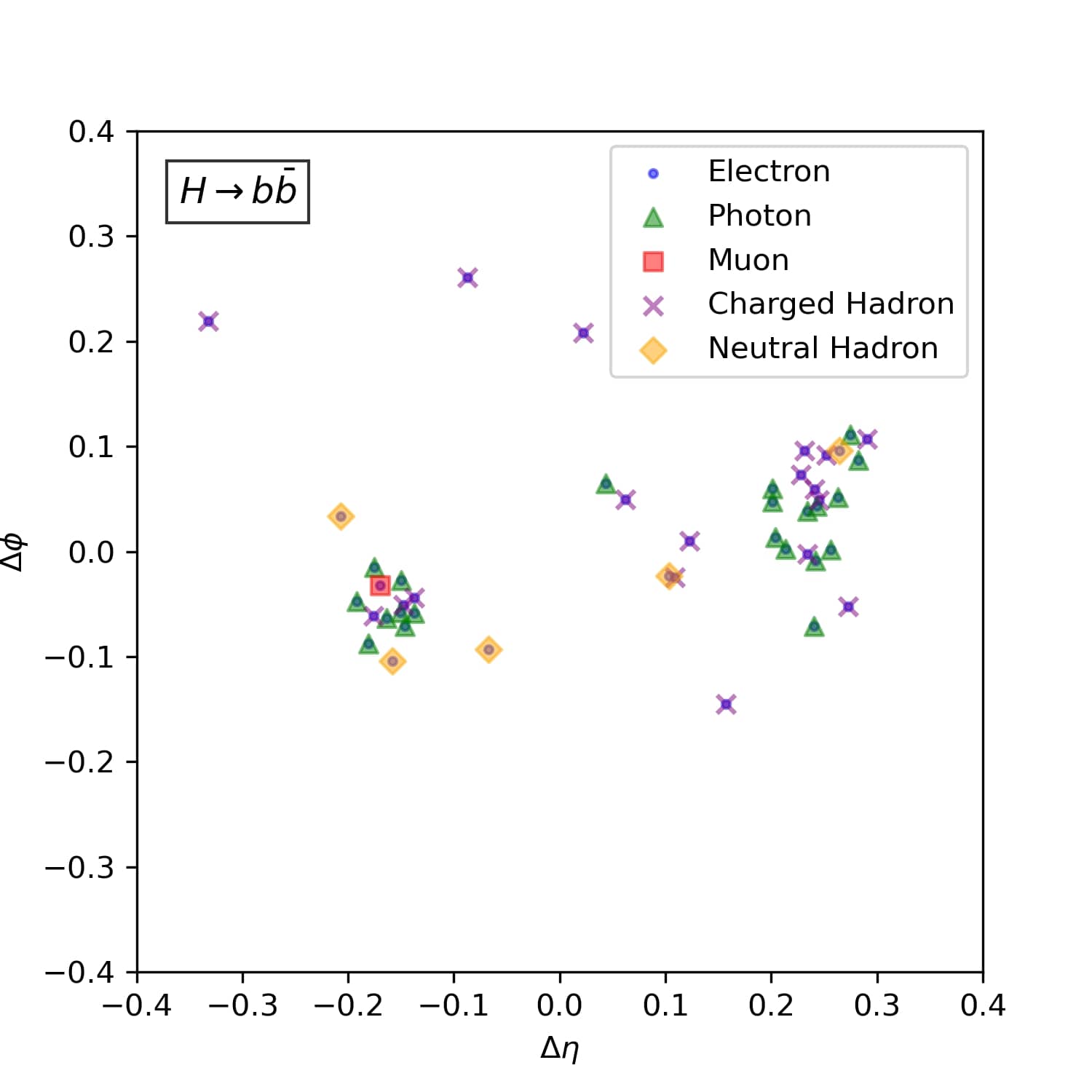}
    \caption{Cloud representation of an example of Higgs jet from JetClass. It shows all particles included in an example from Higgs jet along with the background noise. Higgs decays in to bottom quark anti quark pair which further decays into hadrons that are detected by detectors and are shown here.}
    \label{fig:jet_example}
\end{figure}

JETCLASS was generated by simulating jets in Monte Carlo event generators, using MAD-GRAPH5\_aMC@NLO\cite{alwall_automated_2014}, PYTHIYA\cite{noauthor_introduction_nodate} and DELPHES\cite{the_delphes_3_collaboration_delphes_2014}. MAD-GRAPH\_aMC@NLO simulates the hard processes. Hard processes involve the production of particles after collision and their decay into other particles. Simulation of soft processes that include parton showering and hadronization, is carried out using PYTHIA. In the end, the effects of the detector are simulated by using DLPHES, and we get our final dataset. Important to note that only high-quality jets with 500-1000 GeV transverse momentum and pseudorapidity less than 2 were selected while compiling the dataset and the rest were ignored. Figure \ref{fig:jet_example} shows cloud represention of an example of jet from our dataset.

JETCLASS dataset is the TTree object of the ROOT library with 41 branches. 10 branches are for labels and have boolean values. The rest of the branches are features and correspond to jets or individual particles. The following 3 categories have some most important features for our jet tagging task.

\textbf{Kinematics of particles:} In particle physics, we have a 4-vector ($E, p_x, p_y, p_z$) that includes the energy and momentum of particles and is used to describe the kinematics of particles. 4-vector is directly measured by our detector and is the most fundamental quantity which provides basis for all other kinematics variables. That's why it is a fundamental feature in jet tagging studies.

\textbf{Charge on particles:} Particle charge is crucial in determining that either particle is hadron or photon or electron etc. CMS uses it to identify the particle, and in the dataset, one hot encoding is used to determine this. So we have five variables for each particle type and each of them is hot encoded. For example, to determine if a particle is a photon, we have variable part\_isPhoton and it is hot encoded such that in jet, if a particle is a photon it would have a value of 1, otherwise it would be 0. So we have 5 variable as is\_ChargedHadron, is\_NeutralHadron, is\_Photon, is\_Electron, and is\_Muon for each particle. Recent studies showed that it is important for the classification of decay processes including leptons\cite{qu_particle_2024}.

\textbf{Trajectory features:} This feature is important for jets that involve bottom and charm quarks\cite{cms_collaboration_identification_2020}. This feature was not included in datasets before JetClass. This feature is the impact parameter of particle trajectories which is the smallest distance to the primary vertex value. We have two variables for this, one for longitudinal and one for transverse impact parameters. We also have two extra variables that record the uncertainties in both impact parameters. We have an impact parameter only for charged particles while zero is assigned to the neutral particles.

\textbf{Training set:} In the training dataset, we have 10 M examples of each type of jet. So, we have 100 M jets in total to train our model.

\textbf{Validation set:} A separate dataset of 5 M jets would be used for validating our model during training. This dataset also have equal contributions from each class and has 500 K examples for each type of jet.

\textbf{Test set:} A test data set of 20 M jets (2 M from each jet type) would be used after training the model, for model evaluation purposes.

\section{Model Architecture}
This paper introduces a modified transformer-based architecture called Particle Multi-axis Transformer(ParMAT), a combination of Particle Transformer and Multi-axis Transformer. We split attention into two types: local and global\cite{tu2022maxvit}. Our approach uses layers of block and grid attention in sequence. Block region operates within discrete, non-overlapping windows (small patches on intermediate feature maps) to identify local patterns. In contrast, grid region targets long-range (global) interactions by focusing on a uniformly spaced, sparse grid. The sizes of the windows for both block and grid attentions are adjustable parameters, allowing us to maintain linear computational complexity relative to the size of the input. It serves as a core component facilitating both local and global spatial interactions within a single unit. ParMAT offers increased adaptability and effectiveness compared to full self-attention, showcasing its ability to handle various input lengths with linear complexity. Unlike traditional local attention mechanisms, ParMAT expands model capacity by incorporating a global receptive field. ParMAT is constructed by stacking a particle attention multi-axis block, particle multi-head attention, multi-axis block attention, multi-axis grid attention and class attention mechanism. ParMAT is a modified version of ParT inspired by attention mechanism\cite{qu2024particle}. The model of this architecture is shown in Figure 2(a). We consider $N$ number of particles in a jet. \\\\
\textbf{Input Processing:} ParMAT takes  Particles and Interaction as an input. Particles contains a list of $R$ features for each input particle, which forms a 2-dimensional array of a shape $(N, R)$. Interactions comprises a matrix with $R'$ features for each particle's pair, making a 3-dimensional array $(N, N, R')$. \\\\
\textbf{MLP Projection:} Each set of inputs is passed through a Multi-Layer Perceptron (MLP) to project the $R$ and $R'$ features into $n$ and $n'$-dimensional embeddings, $x_0 \in \textbf{R}^{N \times n}$ and $W \in \textbf{R}^{N \times N \times n'}$ for particles and Interactions respectively, where $x0$ and $W$ are the inputs after an MLP has processed them.
As the order in which particles in a jet are processed does not matter, we do not need any positional encoding. So we feed $x_0$ into multiple Particle Multi-Axis Attention blocks. Each block output is fed into the next block as an input. This produces a series of new embeddings\cite{touvron2021going}.
\begin{figure*}[h] 
    \centering
    \includegraphics[width=\linewidth]{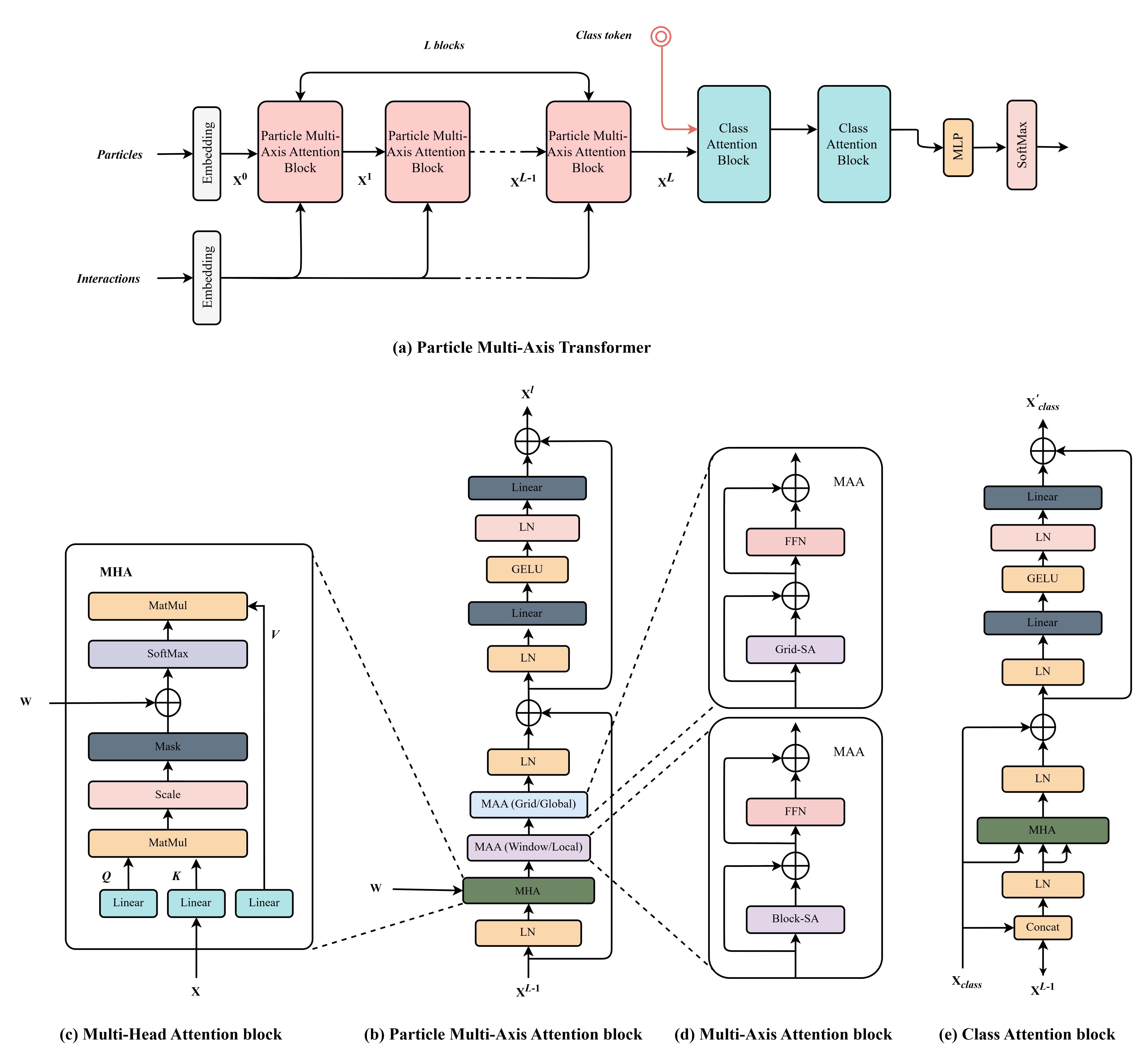}
    \captionsetup{justification=centering}
    \caption{The model diagram of (a) Particle Multi-Axis Transformer, (b) Particle Multi-Axis Attention block, (c) Multi-Head Attention block, (d) Multi-Axis Attention block, and (e) Class Attention Block.}
    \label{fig:image}
\end{figure*} 
\newline\textbf{Interaction features:} Pairwise interaction features are characteristics or properties that emerge from the interaction between pairs of particles. ParMAT is a powerful architecture to find these pairwise interactions, but for the trivial case here we would consider the interaction features extracted from the Energy-momentum 4-vector. For each particle in the Jet Tagging task, we have $p=(E,p_x,p_y,p_z)$. Using the 4-vector $p_m, p_n$ for two particles $m, n$, we calculate 4 features: 

\begin{equation}
\begin{aligned}
\Delta &=  {\left( (y_m - y_n)^2 + (\phi_m - \phi_n)^2 \right)}^{1/2}, \\
k_T &= \min(p_{T,m}, p_{T,n}) \Delta, \\
z &= \frac{\min(p_{T,m}, p_{T,n})}{p_{T,m} + p_{T,n}}, \\
m^2 &= (E_m + E_n)^2 - \|\mathbf{p}_m + \mathbf{p}_n\|^2
\end{aligned}
\end{equation}

where $y_{m,n}$ is the rapidity (the velocity of a particle in a way that is invariant under the Lorentz transformations along the direction of motion), $\phi_{m,n}$ is the azimuthal angle, $p_{T_m,n}$ is the transverse momentum, and $\mathbf{p}_{m,n}$ is the 3-momentum having norm $\| \mathbf{  } \|$. we now take the logarithm of these variables (ln$\Delta$, ln$k_T$, ln$z$, ln$m^2$) as the interaction feature, as the variable has a long tail distribution \cite{dreyer2021jet}.

\textbf{Particle Multi-Axis attention block:} In Figure 2(b), the P-MAA block consists of two parts. First part consists of a LayerNorm (LN) then a multi-head attention (MHA) module and two Multi-axis attention, MAA (grid/global) and MAA (Window/local) with a LayerNorm (LN) layer again. The the second part includes the 2-layers of Multi-layer Perceptron having the LayerNorm (LN) before it and GELU non-linearity in between. Each layer is connected with the following one. The overall block structure is taken from NormFormer. We modified MHA as P-MHA (particle-Multi head attention), which deals with the pairwise particle interaction. The P-MHA is computed as

\begin{equation}
\text{P-MHA}(Q, K, V ) = \text{SoftMax}(QK^T/({d
})^{1/2} + W)V\hspace{0.5cm}
\end{equation}

where, Q, K, and V are the query, key, and value. These are the linear projections of the particle embedding in our case $x^L$. We modify the pre-softmax attention weights by adding the interaction matrix W. This increases the expressiveness of the attention mechanism by enabling P-MHA to alter the dot-product attention weights and integrate particle interaction information.

To make our model more advanced and efficient, we added an extra layer of Multi-Axis attention, which is also called relative attention. We can calculate P-MAA using: 

\begin{equation}
\text{P-MAA}(Q, K, V) = \text{SoftMax}(QK^T/({d})^{1/2} + W)V\hspace{0.5cm}
\end{equation}

\textbf{Class Attention Block:} The class attention block as illustrated in figure 2(b) has almost similar architecture as the particle multi-axis attention block. In this block, we used simple MHA and calculate attention between particles and Global class token $x_{class}$. MHA inputs are: 
 
\begin{equation}
\begin{aligned}
    Q = J_{qx_{class}} + b_q, \\
    K = J_{kz} + b_k, \\
    V = J_{vz} + b_v.
\end{aligned}
\end{equation}

where $J_q$, $J_k$, and $J_v$ are the weight matrices that the model learns for creating the query, key, and value vectors. $x_{class}$ and $z$ are the input vectors. Here $z$ is the product of $x_{class}.x^L$.$b_q$, $b_k$, and $b_v$ are the bias terms.

\textbf{Model Implementation:} We implement our ParMAT model in PyTorch\cite{PaszkeGMLBCKLGA19}. Particularly, P-MHA and P-MAA is implemented using the PyTorch. The ParMAT model includes 8 P-MAA blocks and 2 class attention blocks. We provide interaction matrix $W$ as the attention mask input. It utilizes a 128-dimensional particle embedding, generated from the input particle features through a 3-layer MLP with (128, 512, 128) nodes in each layer, utilizing GELU non-linearity, and LN for normalization. The interaction input features are processed through a 4-layer point wise 1D convolution with channel configurations (64, 64, 64, 8), GELU non-linearity, and batch normalization to create an 8-dimensional interaction matrix. The P-MHA (MHA) in the particle (class) attention blocks have 8 heads, each with a query dimension of 16 and an expansion factor of 4 for the MLP.
We used a dropout (a regularization technique used in neural networks to prevent overfitting) of 0.1 for all P-MAA blocks and no dropout for the class attention block. 

\section{Experiment}

We perform experiments using the new JETCLASS dataset, a widely adopted benchmark for jet tagging tasks in high-energy physics research. The dataset comprises simulated particle collision events, with each event containing detailed information about the particles produced in the collision, including their momentum and energy. The results of our experiment are summarized in Section 7, where we also compare the performance to previous leading methods.

\textbf{Methodology:} In the experiments performed using the JETCLASS dataset, we employ the entire spectrum of particle characteristics as our inputs, covering kinematics, particle identification, and trajectory displacement. These comprise a comprehensive list of 17 features for each particle. Furthermore, we incorporate the four interaction features introduced for the ParMAT. The training process is executed on the entirety of the training set, which consists of 100M jets. We utilized the Ranger optimizer, incorporating a weight decay of 10{-4}. The initial learning rate of 0.001 schedule followed a flat cosine curve with a warm-up period of 50,000 steps. 1024 was selected as the batch size. We implemented it on an NVIDIA DGX system and utilized multiple GPUs (0, 1, 2, and 3) to expedite training. Following training, we evaluated the trained ParMAT model on a separate validation set from the JetClass dataset. 

\textbf{Baselines:} We assess the performance of PARMAT against four baseline models: ParT\cite{qu2022particle}, PFN\cite{komiske2019energy} utilizing Deep Sets\cite{zaheer2017deep}, P-CNN, as employed in the DeepAK8 algorithm of the CMS experiment\cite{drugakov2020identification}, and ParticleNet\cite{qu2020jet}, an advanced architecture derived from DGCNN\cite{moreno2020jedi}, as presented by Qu \& Gouskos (2020). The results documented in the ParT paper for all models trained comprehensively on the JETCLASS dataset are employed for comparison.ParticleNet utilizes an established PyTorch implementation, whereas, for PFN and P-CNN, the ParT paper reconstructed them in PyTorch to guarantee the reproduction of published outcomes. The optimizer and learning rate schedule are consistent with those used in the training of ParMAT.The optimal (Learning rate, batch size) combinations are (0.0010, 1024) for ParMAT, (0.0010, 512) for ParT, (0.010, 512) for ParticleNet, and (0.020, 4096) for PFN and P-CNN.

\section{Evaluation Metrices}
We recommend using key performance indicators to fully assess the performance of ParMAT on the detailed JETCLASS dataset. Due to the fact that jet tagging in this dataset is a classification that includes multiple classes, We advise using two widely used metrics: area under the receiver operating characteristic curve and accuracy. These measurements do a good job of evaluating overall efficacy. Furthermore, We suggest that background rejection which is the opposite of false positive rate should be taken into an account at a specific signal optimization, denoted by the true positive rate also written as TPR set at x\%.

\begin{equation}
  Rej_{xy\%} = 1 / \text{FPR at TPR = x\%}
\end{equation}

We apply this equation to all kinds of signal jets. Quark-Gluon jets are, by default, categorized as background jets, following standard procedure in the majority of LHC data analysis. The remaining nine jet types are all regarded as signal jets. However, for $H \rightarrow l\nu q q'$($ t \rightarrow b l \nu $), this efficiency is increased to 99\% (or 99.5\%) as these types exhibit more distinctive characteristics and are more readily distinguishable from q/g jets. As the Rejection metric concentrates on two classes only – the background class (b) and the signal class (s), the true positive rate and the false positive rate are assessed by using dual class scoring method.

\begin{equation}
  \text{Score}_{svsb}\equiv \frac{\text{Score}(s)}{\text{Score}(s) + \text{Score}(b)}
\end{equation}

In the context of jet tagging, separating signal jets (class s) from background jets (class b) is crucial. This equation, inspired by the methodologies of the CMS experiment at the LHC, utilizes softmax outputs (Score(s) and Score(b)) to achieve optimal discrimination between the two classes. Background rejection is challenging metric in jet tagging. This is due to significant improvement in background rejection directly translates to a substantial increase in the LHC's discovery potential. For example, doubling the background rejection can result in a 40\% increase in discovery potential, which is comparable to either doubling the dataset size or operating the LHC for twice the duration\footnote{The ROC AUC function available in scikit-learn library may be used to calculate the area under the curve by passing in the arguments, average equals to 'macro' and multi-class equals to 'ovo'.}.

\section{Results}

Evaluation metrics described in Section 6 are used to determine the performance of our proposed model on JETCLASS dataset, with the outcomes compiled in Table 1.This new proposed architecture sets a new benchmark by achieving the highest performance of 86.32\%, surpassing all others. ParMAT performs with higher accuracy than Particle transformer.ParMAT significantly outperform the earlier ParT,ParticleNet and P-CNN models on this comprehensive dataset, showing background rejection differences that may exceed an order of magnitude. This significant enhancement in the Particle Multi-Axis Transformer is projected to significantly enhance the discovery capabilities for physics research at the LHC.

\begin{table*}[h]
    \centering
    \scriptsize
    \caption{The performance of our Particle Multi-axis transformer is evaluated and contrasted with the results obtained by Qu et al. in their 2022 study\cite{qu2022particle},where they employed the particle transformer for jet tagging. Our model’s performance is also compared with Particle flow network(Komsike et al.,2019b)\cite{komiske2019energy},P-CNN, as reported by the CMS Collaboration (2020b)\cite{drugakov2020identification} and ParticleNet by Qu and Gouskos,2020\cite{qu2020jet}. ParMAT model has demonstrated higher performance using the JETCLASS dataset.}
    \begin{tabularx}{\textwidth}{|l|*{2}{X|}*{9}{X|}}
        \hline
        \textbf{Model} & \textbf{Accuracy (\%)}& \textbf{AUC  (\%)} & \textbf{Z→qq Rej50\%} & \textbf{t→bqq Rej50\%} & \textbf{t→blv Rej99.5\%} & \textbf{W→qq Rej50\%} & \textbf{H→bb Rej50\%} & \textbf{H→cc Rej50\%} & \textbf{H→gg Rej50\%} & \textbf{H→4q Rej50\%} & \textbf{H→lvqq Rej99\%} \\
        \hline
        PFN & 77.20 & 97.10 & 159.0 & 797.0 & 721.0 & 189.0 & 2924.0 & 841.0 & 75.0 & 198.0 & 265.0 \\
        \hline
        PCNN & 80.90 & 97.80 & 204.0 & 2907.0 & 2304.0 & 241.0 & 4890.0 & 1276.0 & 88.0 & 474.0 & 947.0 \\
        \hline
        ParticleNet & 84.40 & 98.40 & 283.0 & 10526.0 & 11173.0 & 347.0 & 7634.0 & 2475.0 & 104.0 & 954.0 & 3339.0 \\
        \hline
        ParT(Plain) & 84.90 & 98.50 & 311.0 & 17699.0 & 12987.0 & 384.0 & 9569.0 & 2911.0 & 112.0 & 1185.0 & 3868.0 \\
        \hline
        ParT & 86.10 & 98.70 & 402.0 & 32787.0 & 15873.0 & 543.0 & 10638.0 & 4149.0 & 123.0 & 1864.0 & 5479.0 \\
        \hline
        ParMAT & 86.23 & 98.75 & 116.3 & 32906.7 & 24793.3 & 160.0 & 4484.3 & 1509.9 & 227.4 & 503.1 & 4211.5 \\
        \hline
    \end{tabularx}
    \label{Table 1}
\end{table*}
We are plotting two graphs here: one with the x-axis representing time (epochs) and the y-axis representing the model's accuracy. These graphs visualizes the performance of Particle Multi Axis Transformer models on JETCLASS dataset. Figure 3 illustrates the plot of training accuracy versus epochs, while Figure 4 displays the graph of validation accuracy versus time (epochs). As training progresses, the graphs illustrate how the model's performance improves. The y-axis ranges from 0 to 100\% but accuracy levels off at around 85\% ,indicating that its capacity may not be sufficient to learn beyond that point. This could also suggest that the model might be over-fitting.By analyzing the curves, it's apparent that ParMAT consistently yields higher accuracy suggesting its potential efficacy in particle collider studies for distinguishing between different types of jets.
\begin{figure}[H]
    \centering   
    \subfloat[]
    {\includegraphics[width=0.45\linewidth]{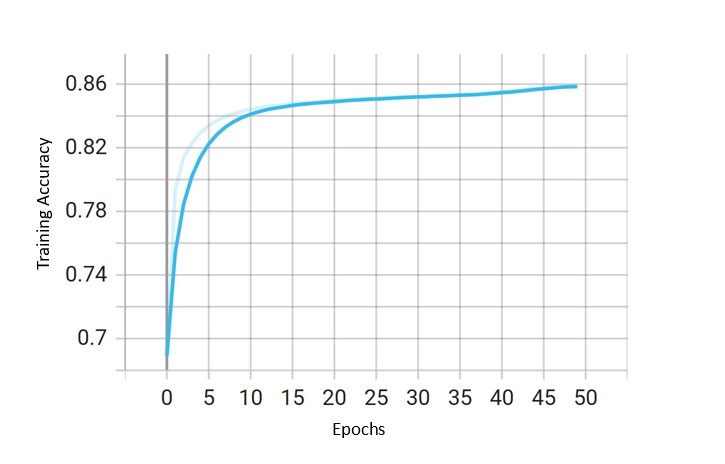}}
    \hspace{0.02\linewidth} 
    \subfloat[]
    {\includegraphics[width=0.45\linewidth]{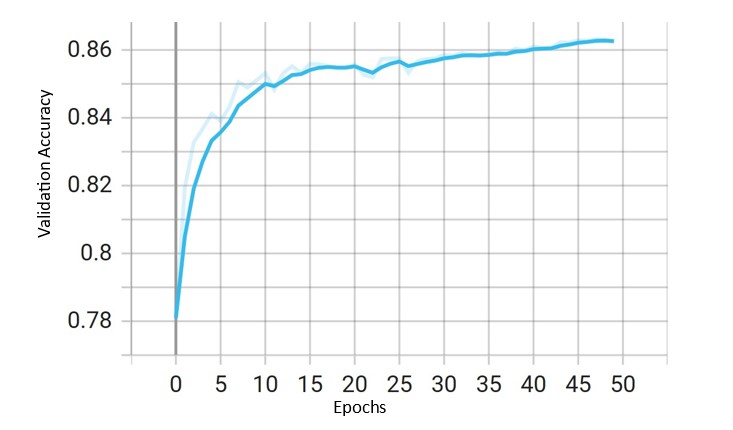}}  
    \caption{(a) Training Accuracy vs Epochs and (b) Validation Accuracy vs Epochs.}
\end{figure}
Tagging performance shows notable variations across different signal types. Exceptional discrimination against background quark/gluon jets is observed with  $ t \rightarrow b l \nu $ and $ H \rightarrow l \nu q q' $ signals. Employing latest ParMAT architecture allows for almost complete detection of these signals, achieving over 99\% efficiency with minimal contamination from background jets. This breakthrough, utilizing jet types previously not being discovered for tagging at the LHC, paves the way for new opportunities in this domain.

\textbf{Effect of training dataset Size on ParT Performance}: To assess how the dataset size influences jet tagging performance, we conducted further training sessions and replicated the results of the ParT model using incomplete data\cite{karwowska2024particle}. The original configuration of the particle transformer included 8 attention blocks with 8 attention heads each, utilizing the GeLU activation function with a learning rate (LR) of 0.001. The Ranger optimizer was used to enhance the model's performance. Batch size was set to 512. In our initial experiment, preserving this original setup, we achieved an accuracy of 85.43\%. However, when the training dataset size was reduced to 1\%, there was a 0.67\% decrease in accuracy. In the second experiment, we increased the number of attention blocks and heads to 16 each and switched to the Mish activation function, while continuing to use only 1\% of the dataset. This modification resulted in an improved accuracy of 86.15\%, with the same optimizer, learning rate, and batch size.

\textbf{Effect of training dataset Size on ParMAT Performance}:To investigate the impact of reduced class-specific data on overall performance, we retrained the model over 20 epochs utilising only 50\% of the original JETCLASS dataset in our most recent training iteration. 8 attention heads and the same number of layers made up the architecture's configuration. We employed the Mish activation function due to its proven effectiveness in handling non-linearities. To reduce overfitting, a dropout rate of 0.20 was used in the training, and the Ranger optimizer was used due to its effectiveness in convergent to minima. The batch size was set to 1024.The model's accuracy decreased noticeably to 85.86\% in spite of these configurations.This reduction may be attributed to the decreased volume of training data for the JETCLASS, suggesting that further adjustments in data handling or model parameters might be necessary to restore or enhance performance.
\section{Conclusion}
We proposed Multi-axis particle transformer in this paper for jet tagging task on particle collision data from CMS detector. Our model performed well on JETCLASS dataset by achieving a reasonable accuracy for classifying our signals into 10 classes and surpassed many previous ML models proposed for jet tagging including the four baselines models. The AUC of 98.76\% reflects the strong discriminatory power of our model and its robustness. It also shows the model's ability to correctly identify our signals while minimizing the false positives. It also suggest the ability of our model for jet tagging task where precise classification is crucial and false positives can lead to significant drawbacks. Ranger optimizer performed better than Adam for training this model (Complete details about training parameters have been covered in section 5: Experiment). Background rejection matrix shows that our model significantly improved on background noise rejection for both decay modes of top quarks $(t\xrightarrow{}bqq'$ \& $t \xrightarrow{}blv)$ and one decay mode of Higgs$(H\xrightarrow{}gg)$. This improvement resulted from integration of multi axis attention block which uses multi axis attention mechanism to explore relations between different particles and jets by considering multiple axis at a time. Results support the fact that transformers can indeed be a good solution for classification tasks in particle physics where we have large amounts of data generated by detectors. In the future, jet tagging in particle physics experiments is expected to evolve significantly\cite{fermi_studies_1955}. Researchers can harness particle level information, using individual particle data within jets, including track information and calorimeter shower shapes, to enhance tagging accuracy. Additionally, the global event context can play a crucial role, with algorithms incorporating event-wide data such as missing energy and jet multiplicity to refine tagging strategies. Time series analysis can be employed to distinguish jets in complex events with multiple proton-proton collisions, while calorimeter substructure techniques can be refined to exploit the intricate details of jet formation for improved tagging power.
\section*{Acknowledgements}
We express our gratitude to the Pattern Recognition Lab (PR Lab) and the Pakistan Institute of Engineering and Applied Sciences (PIEAS) for offering the essential computational facilities and fostering a conducive research atmosphere.

\section*{Declarations}

\subsection*{Funding/Competing interests}
The authors affirm that they have no known conflicts of interest, whether financial or personal, that could have influenced the research presented in this article.

\bibliographystyle{unsrt}  
\bibliography{references}

\end{document}